\documentclass[aps,twocolumn,prb,showpacs,floatfix]{revtex4}

\usepackage{graphicx}
\usepackage{amsmath}
\usepackage{epsfig}
\usepackage{dcolumn}
\usepackage{bm}

\begin{document}

\title{Large Magnetoresistance 
in a Manganite Spin-Tunnel-Junction \\
Using LaMnO$_3$ as Insulating Barrier} 

\author{S. Yunoki, and E. Dagotto}

\affiliation{
Department of Physics and Astronomy, The University of Tennessee, Knoxville, 
Tennessee 37996, USA, 
and Materials Science and Technology Division, Oak Ridge National Laboratory, 
Oak Ridge, Tennessee 32831, USA.}

\author{S. Costamagna and J. A. Riera}
\affiliation{Instituto de F\'{\i}sica Rosario, Consejo Nacional de
Investigaciones Cient\'{\i}ficas y T\'ecnicas,\\
Universidad Nacional de Rosario, Rosario, Argentina
}

\date{\today}

\begin{abstract}

A spin-tunnel-junction based on manganites, with 
La$_{1-x}$Sr$_x$MnO$_3$ (LSMO) as ferromagnetic metallic electrodes 
and the undoped parent compound LaMnO$_3$ (LMO) as insulating barrier, 
is here theoretically discussed using double exchange model Hamiltonians 
and numerical 
techniques. For an even number of LMO layers, the ground state is shown 
to have anti-parallel LSMO magnetic moments. This highly resistive, but 
fragile, state is easily destabilized by small magnetic fields, which 
orient the LSMO moments in the direction of the field. The 
magnetoresistance associated with this transition is very large, 
according to Monte Carlo and Density Matrix Renormalization 
Group studies. 
The influence of temperature, the case of 
an odd number of LMO layers, and the differences between LMO and 
SrTiO$_3$ as barriers are also addressed. General trends are discussed.

\end{abstract}


\maketitle

\section{Introduction}\label{intro}

The study of strongly correlated electronic systems (SCES) continues 
attracting the attention of the Condensed Matter community. 
These materials present complex phase diagrams that illustrates 
the competition 
which exists among phases with very different physical properties, such as 
$d$-wave superconductivity, antiferro- and ferro-magnetic order, 
charge- and orbital-order, multiferroic behavior, and several others. 
Moreover, this complexity and phase competition lead to self-organized 
nano-scale inhomogeneities 
which are believed to generate giant responses, 
as in the famous colossal magnetoresistance (CMR) effect of the Mn-oxides 
known as manganites.\cite{dagotto-CMR}

Recently, a new procedure to study oxide SCES has been proposed. 
It involves the artificial creation of oxide multilayers with atomic-scale 
accuracy at the interface, via the use of techniques such as pulsed-laser 
deposition.\cite{multilayers} Potentially, these structures can have 
properties very different from those of the building blocks. One of the 
reasons for this expectation is that a transfer of charge could occur 
between the constituents leading, for example, to the stabilization of a 
metal at the interface between two insulators.\cite{multilayers} 
The creation of novel two-dimensional states, as well as the possible 
applications of oxide multilayers in the growing field of oxide electronics, 
has given considerable momentum to these investigations.

\subsection{Spin-Tunnel-Junctions}\label{intro_stj}
The development of the above mentioned accurate experimental techniques 
for the construction of oxide multilayers with atomic precision can have 
implications in the study of 
spin-tunnel-junctions,\cite{junctions,junctions-reviews} introducing 
a better control of their properties. These structures consist of two 
ferromagnetic (FM) metallic electrodes, separated by a thin insulating 
barrier. 
The resistance of this device depends on the relative orientation of the 
electrodes' magnetizations. The tunneling magnetoresistance (TMR) is 
usually defined via the difference in resistances between the anti-parallel 
and parallel arrangements of the electrodes' magnetic moments. Half-metals, 
such as La$_{1-x}$Sr$_x$MnO$_3$ (LSMO), with an intrinsic nearly full 
magnetization, are ideal for these devices.\cite{half-metal}

In the context of magnetic tunnel junctions, very interesting results were 
reported by Bowen {\it et al.}~\cite{bowen} using a LSMO/STO/LSMO trilayer. 
The Sr concentration was 1/3 for La$_{1-x}$Sr$_x$MnO$_3$ (LSMO), 
and STO represents the insulator 
SrTiO$_3$. A huge TMR ratio of more than 1800\% was 
observed at very low temperatures 4~K, showing the advantages of using 
half-metallic LSMO as a ferromagnetic electrode in the junctions.
However, in the same investigations it was reported that the large TMR 
survived only up to $\sim$270~K, lower than the Curie temperature of 
LSMO($x$=1/3), which is $\sim$370~K. It was argued that the deterioration 
of the ferromagnetism near the LSMO/STO interface (``dead layer'') could be 
causing this TMR reduction. Later, Yamada {\it et al.}~\cite{yamada} 
addressed this problem by comparing the STO/LSMO interface with others, 
such as LAO/LSMO or STO/LMO/LSMO, where LAO stands for LaAlO$_3$ and LMO 
for LaMnO$_3$. Those authors found that the magnetic behavior of LAO/LSMO 
and STO/LMO/LSMO are much better than STO/LSMO in the sense that no 
dead-layer was found, opening a new path toward LSMO-based TMR junctions 
operating at room temperature.

\subsection{Proposed Main Idea}\label{idea}
\begin{figure}[hbt]
\includegraphics[clip=true,width=7.5cm,angle=-0]{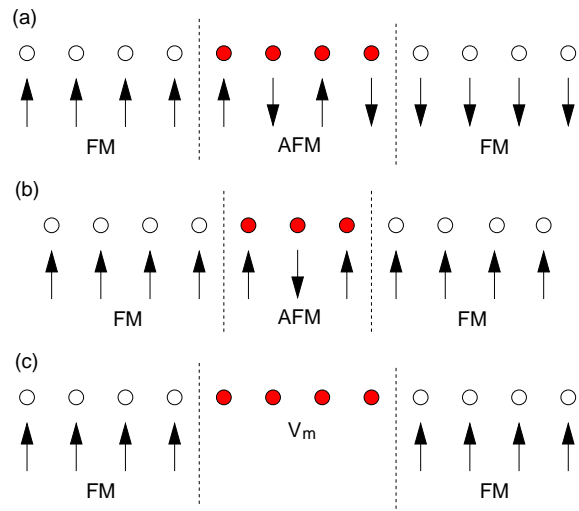}
\begin{center}
\caption{
(Color online) 
Schematic representation of a one-dimensional (1D) double exchange (DE) 
model for 
a LSMO/LMO/LSMO trilayer with even (a) and odd (b) numbers of LMO sites. 
Here, LMO and LSMO represent regions where the positive background 
charge densities correspond to an antiferromagnetic (AFM) insulator 
and ferromagnetic (FM) metal, respectively, when in bulk form. 
Circles stand for lattice sites where electrons can move, and 
arrows indicate localized spins. 
The trilayer with even (odd) number of LMO sites gives rise to 
anti-parallel (parallel) alignment of the magnetic moments in the left 
and right 
ends of LSMO. (c) Schematic representation of a 1D model for a LSMO/STO/LSMO 
trilayer. Here, LSMO is described by the DE model as before, while STO is 
modeled by a tight-banding model with no localized spins. 
$V_{\rm m}$ introduced in the central STO region is a band off-set 
site-potential mimicking the work function 
difference between LSMO and STO. More details can be found in the text. 
}
\label{spin}
\end{center}
\end{figure}

In this paper, an alternative setup is proposed for a manganite trilayer 
system which is expected to have a very large magnetoresistance (MR) at 
low temperatures, at least according to modeling calculations reported below. 
The proposed geometry, and main idea behind its performance, is presented 
in Fig.~\ref{spin}. The new system is made out entirely of manganite 
materials, with different hole doping concentrations.\cite{jo} 
Using all manganites may help in the interfacial contact between the 
components due to their similar lattice spacings. More specifically, 
in Fig.~\ref{spin} (a) a trilayer system is represented. It contains two 
hole-doped Mn-oxides, such as LSMO, and a central region made out of the 
undoped parent compound LMO. It is well known that LSMO is ferromagnetic 
and half-metallic for sufficiently large hole doping, while LMO is an A-type 
antiferromagnetic (AF) insulator.~\cite{dagotto-CMR} The arrows in 
Fig.~\ref{spin} (a) represent schematically the expected spin orientations 
in a one-dimensional arrangement, for simplicity. The most interesting 
detail of Fig.~\ref{spin} (a) is the relative orientation of the spins 
between the metallic leads. For an $even$ number of layers in the central 
LMO region, the ferromagnetic moments of the leads are $anti$-$parallel$. 
This is expected to cause a very large resistance at low temperatures, 
since the carriers moving from one lead to the other not only must tunnel 
through the central insulating barrier, but in addition the spin species 
which can travel in one lead is blocked by the other. However, the 
anti-parallel configuration, which is mediated by the central region, is 
not strongly pinned in this arrangement: it is only a $weak$ 
antiferromagnetic effective interaction which produces the ground state 
with anti-parallel leads' moments. Thus, relatively small magnetic fields 
can render the ferromagnetic moments parallel, substantially reducing the 
resistance. All these intuitive ideas will be substantiated via model 
calculations, described below.

For an $odd$ number of layers in the central LMO region, the expected spin 
arrangement is shown in Fig.~\ref{spin} (b). In this case the magnetic 
moments of the leads will be parallel to one another, and the resistance 
will not be as large as for the configuration shown in Fig.~\ref{spin} (a). 
However, the LMO region still provided a tunneling barrier, and the 
performance of this case will be shown below to be 
quite acceptable, at least within modeling calculations.

Finally, Fig.~\ref{spin}(c) contains a crude representation of a more 
standard trilayer device.~\cite{bowen} While the leads are still representing 
LSMO, the central region is now a band insulator, as it occurs in the much 
employed case of STO. The main focus of our effort will be the cases shown in 
Fig.~\ref{spin} (a) and (b), but some results for the setup shown in 
Fig.~\ref{spin} (c) will also be discussed for completeness, and to clarify 
qualitative trends.

\subsection{Main approximations}\label{main_appr}

Before proceeding to the presentation of the results, some of the 
theoretical approximations used must be clearly expressed for the benefit 
of the reader. With this paper, our main intention is to motivate experimental 
groups to consider the materials and setups proposed in our study, 
involving a magnetically active manganite barrier, as opposed 
to a magnetically-inert band-insulator, as the widely used STO. However, 
it must be clearly stated that our calculations are qualitative at best and 
should be considered only as a guidance to understand the intuitive picture 
presented here.

A variety of effects are not taken into account in this investigation 
(and actually these effects cannot be taken into account accurately within 
the current status of numerical simulations and Hamiltonian modeling). 
{\it (1)} For instance, the lattice and orbital reconstructions are not 
incorporated, but only the electronic reconstruction is considered. In other 
words, the atomic positions are considered to be rigid here. It would be 
the task of sophisticated {\it ab-initio} simulations to consider how the 
lattice relaxes at the interface between LMO and LSMO, and such future 
calculations are certainly strongly encouraged. 
{\it (2)} A second subject which is not addressed in this work is the issue 
of the infamous ``dead layer'' at the barrier/electrode interface, already 
briefly described in Sec.~\ref{intro_stj}. Dead layers affect the performance 
of several trilayer devices. We have attempted to mimic this dead layer 
altering by hand the chemical potential in the vicinity of the interface, 
but the results were not sufficiently satisfactory to be described in this 
paper. Thus, this problem is left for future efforts. 
{\it (3)} Another topic which is only briefly discussed is the influence of 
anisotropies: in most of the simulations below there is no ``easy'' axis 
introduced for the magnetization orientation. While for manganites in bulk 
form this is a reasonable assumption, in thin films it is known that in-plane 
magnetic anisotropies are induced by epitaxial strain.\cite{mathews} 
These anisotropies are the reason behind the abrupt changes in resistances 
observed when varying magnetic fields in magnetic tunnel junctions 
(see, for instance, Fig.~2 in Ref.~\onlinecite{bowen}). While in our 
numerical studies reported below the simulations are carried out mainly with 
isotropic Heisenberg spins, a more proper analysis would have needed 
anisotropic terms, rendering these spins Ising like. 
{\it (4)} A related topic involves the experimentally observed differences 
between the upper and lower electrodes in a trilayer junction, since the 
area of the bottom electrode is typically much wider than that of the upper 
one.\cite{ishii} This stabilizes anti-parallel orientations of the 
electrodes' magnetic moments at some magnetic fields. In our simulations, 
both electrodes are perfectly equivalent. However, for the reasons already 
mentioned in the previous subsection, we do stabilize a similar 
anti-parallel magnetic moment arrangement via the use of LMO as barrier, 
with an even number of sites and in zero external magnetic field. 
{\it (5)} Finally, a practical assumption in our investigations is the 
focus on a one-dimensional (1D) spin arrangement, described by the realistic 
double-exchange model for manganites (albeit restricted to just one 
orbital). The restriction to a 1D configuration is needed for the numerical 
studies to be accurate; higher dimensional arrangements would have increased 
so much the CPU time that a careful analysis would have been impossible. 
Note that contrary to manganite bulk studies,\cite{dagotto-CMR} 
where the analysis of small two-dimensional clusters is possible, here we 
will be carrying out an iterative loop to solve Poisson's equation 
which regulates the charge transfer between materials 
(see Sec.\ref{model-methods}). At each step of the iterative process, 
an {\it entire} Monte Carlo (MC) process or Density Matrix Renormalization 
Group (DMRG) sweep is carried out (see details below), thus increasing 
substantially the computer requirements. Then, the restriction to a 1D 
geometry is caused by the CPU resources available. However, in the 
description of results below, we have focus on qualitative aspects which 
are expected to be robust, and we strongly believe that they will survive 
the increase in dimensionality. For example, the large magnetoresistance 
at low temperatures of the LSMO/LMO/LSMO setup, with an even number of 
layers for LMO, is believed to occur in any dimension of interest.

\subsection{Organization}
The organization of the paper is the following. In Sec.~\ref{model-methods},  
the model Hamiltonians and numerical methods are described. The focus is on 
Monte Carlo and DMRG techniques. The main results are presented 
in Sec.~\ref{T=0}, which correspond to the arrangements schematically 
described in Fig.~\ref{spin} (a) and (b), showing that the magnetoresistance 
is large in these setups. Both classical and quantum localized $t_{\rm 2g}$ 
spins are used. In Sec.~\ref{Tnonzero}, the influence of temperature is 
analyzed. As observed in some experiments,~\cite{bowen} it is found that the 
large MR effect quickly deteriorates with increasing temperature. 
The influence of anisotropies is also studied as a possible cure 
to this problem. In Sec.~\ref{STO}, for completeness, results of the 
modeling of a trilayer involving a band insulator (such as STO) as the 
barrier, instead of LMO, are reported. 
Conclusions are given in Sec.~\ref{conclusions}. 


\section{Model and methods}
\label{model-methods}

\subsection{Model Hamiltonian and MC methods}

To crudely model LSMO/LMO/LSMO trilayers (Fig.~\ref{spin}), the one-orbital
double exchange (DE) model on a 1D lattice will be used:
\begin{eqnarray}\label{model}
H&=&-t\sum_{\langle i,j \rangle}\sum_\sigma\left(c_{i,\sigma}^\dag c_{j,\sigma}
+{\rm  H.c.}\right) + \sum_i \phi(i) n_i \nonumber\\
&&-J_{\rm H}\sum_i {\cal{\bf s}}_i\cdot{\bf S}_i-\sum_i{\bf h}_{\rm ext}\cdot{\bf M}_i,
\end{eqnarray}
where $c_{i,\sigma}^\dag$ is the creation operator of an electron at site $i$ 
with spin $\sigma(=\uparrow,\downarrow)$. 
The first summation of $\langle i,j \rangle$ runs over nearest neighbor 
pairs of sites $i$ and $j$. 
The number operator is $n_i=\sum_{\sigma}c_{i,\sigma}^\dag c_{i,\sigma}$,
$\phi(i)$ is the electrostatic potential (discussed below), and 
the spin operator of the electron is 
${\bf s}_i=\sum_{\alpha,\beta}c_{i,\alpha}^\dag
\left({\vec\sigma}\right)_{\alpha\beta} c_{i,\beta}$
(here ${\vec\sigma}=(\sigma_x,\sigma_y,\sigma_z)$: Pauli matrices).
${\bf S}_i$ is the classical spin, widely used to represent the localized 
$t_{\rm 2g}$ spins, with $|{\bf S}_i|$=$1$. $J_{\rm H}$ is the Hund's rule 
coupling, ${\bf h}_{\rm ext}$ is an external magnetic field, and 
${\bf M}_i$=${\frac{1}{2}}({\bf s}_i+3{\bf S}_i)$ is the total magnetic 
moment at site $i$. Hereafter, $t$ is set to be 1 as an energy unit. 
The number of sites in the left, central, and right regions of the system 
is denoted by $L^{(\rm L)}$, $L^{(\rm C)}$, $L^{(\rm R)}$, respectively.

To study the electronic properties of hetero-structured systems, it is 
crucial to include the cation ions in the model 
(to consider the charge neutrality condition)  and take 
into account long-range Coulomb interactions between electrons and cation 
ions. In this study, the long-range Coulomb interactions are considered 
within the Hartree approximation through Poisson's equation:~\cite{datta}
\begin{equation}\label{poisson}
\nabla^2\phi(i) = -\alpha\left[\langle n_i\rangle-n_+(i)\right],
\end{equation}
where $n_+(i)$ is the positive background charge mimicking the cation ions. 
Here in this paper, $n_+(i)$=$n^{(\Gamma)}$ is set to be uniform within each 
layer [$\Gamma$=${\rm L}$ (left region), C (central region), and 
R (right region)], with a value determined by the charge neutrality 
condition. The parameter $\alpha$=$e^2/\varepsilon a$ ($e$: electronic 
charge, $\varepsilon$: dielectric constant, $a$: lattice constant) 
is the strength of the Coulomb interactions, considered as a free parameter 
in the model. 
To solve Eq.~(\ref{poisson}), the symmetric discretization of 
Poisson's equation is used, which in 1D becomes 
$\nabla^2\phi(i)$=$\phi(i+1)-2\phi(i)+\phi(i-1)$, 
with boundary conditions $\phi(i)$=$0$ for sites outside of the system.

In the following sections, results for zero temperature ($T$) as well as 
finite temperatures are reported. For the finite temperature calculations, 
the standard grand-canonical Monte Carlo simulation is used,~\cite{yunoki} 
with the 
chemical potential $\mu$ adjusted such that the total number of electrons 
are equal to the total background positive charge $\sum_i n_+(i)$. For the 
zero temperature calculations, a full optimization of $\{{\bf S}_i\}$ using 
the Broyden-Fletcher-Goldfarb-Shanno method~\cite{NR} is performed 
for a fixed number $N$ of electrons, i.e., $N=\sum_i n_+(i)$ 
(canonical ensemble). We have found that the Monte Carlo simulations 
at very low temperatures produce almost identical results as those 
calculated using the zero temperature canonical method, although at 
a considerably larger cost in CPU time. Thus, at $T$=0 it is advantageous 
to use the optimization method. Note that due to the need to solve 
Poisson's equation iteratively, a Monte Carlo simulation or optimization 
procedure has to be carried out at each step, increasing substantially the 
computer time as compared with more standard simulations of bulk systems. 
The calculation of the conductances is carried out using the Landauer 
formalism,~\cite{verges} as extensively explained in previous 
reports.~\cite{dagotto-CMR,dagotto-book}

In the real LSMO/STO/LSMO trilayer systems, which will also be briefly 
modeled in this manuscript, it is important to notice that there exists band 
offsets (due to work-function differences) between LSMO and STO, which are 
typically of the order of a few eVs,~\cite{sawa} although its 
precise value is difficult to find experimentally.\cite{electron-doped} 
In this paper, this band offset is treated as a parameter described simply by
the addition of a site potential term $V_m\sum_{i\subset{\rm C}}n_i$ to the 
Hamiltonian [Fig.~\ref{spin} (c)]. Moreover, a simple 
tight-binding model without $J_{\rm H}$ in Eq.~(\ref{model}) is used to 
model the STO barrier. Omitting Hubbard-type interactions might be justified 
by the fact that the number of electrons in the central region (STO) is 
very small, as shown later. And for LSMO and LMO, neglecting the Hubbard 
term is justified by the large value of the Hund's coupling which by itself 
prevents double occupancy, as widely discussed before.\cite{dagotto-CMR}

\subsection{Quantum localized spins and DMRG}\label{dmrg2}

Results using quantum spin 1/2 for the localized spins ${\bf S}_i$ in 
Eq.~(\ref{model}) are also presented in this paper, and they are compared 
with results of the classical spin simulations. In real manganites 
$t_{\rm 2g}$ spins are 3/2. However, using spin 1/2 much simplifies 
the computational task due to the reduction in the size of the Hilbert 
space.~\cite{yunoki} 
To study the ground state properties of Eq.~(\ref{model}) with 
${\bf S}_i=1/2$, e.g., the charge density distribution, the 
standard DMRG algorithm,\cite{DMRG} embedded in a self-consistent iterative 
procedure to solve Poisson's equation for the long-range Coulomb potential, 
is used.
                     
More specifically, starting from an initial electrostatic potentials 
$\phi(i)$ ($i=1,\ldots,L$, here $L$ is number of total sites), we 
first make two sweeps for ``warming'' 
before beginning the self-consistent calculations for the long-range Coulomb 
interactions. Then, for the next 10 sweeps, the electrostatic potentials 
$\phi(i)$ are updated by solving Poisson's equation [Eq.~(\ref{poisson})] 
at each sweep. In 
general, the convergence of the self-consistent procedure is very 
slow. Therefore, here we perform an extrapolation of $\phi(i)$ for each 
$i$, using the values calculated during those 10 sweeps, to an infinite 
number of sweeps. By comparing fully self-consistent calculations, we have 
found that generally this extrapolation scheme gives a reasonable results. 
Finally, the obtained potential $\phi(i)$ is plugged back in 
Eq.~(\ref{model}), and 8 more sweeps are 
performed to calculate the ground state. The results reported below are 
obtained by retaining $M$=$350$ states, and the truncation error in the worst 
case is of order $10^{-7}$. All DMRG calculations are done at $T=0$, 
and $J_{\rm H}$=36 is used.~\cite{yunoki}

As a check of our DE model code, we have compared the case of a total 
$z$-projection of the spin $S_z=S_{\rm max}$=$26$, on a lattice of 
$L$=$32$ sites and with a total number of conduction electrons $N=20$, 
with the results of the spinless fermion model which is computed with 
a previously prepared DMRG code for the Hubbard model,~\cite{seba} written 
completely independently from the present DE code. The potential $\phi(i)$ 
added in the spinless fermion calculation is the one extracted from the DE 
run. The differences in energies are found to be of order $10^{-7}$.

To study transport properties, the time-dependent DMRG technique is 
used.\cite{tDMRG} The electrostatic potentials $\phi(i)$ are fixed to the 
ones obtained above, and a small bias potential $\Delta V$=$0.01$ is 
applied at time $t$=$0$, which triggers a time-evolution of the 
system.~\cite{alhassanieh}  The bias potential is applied only on 
a few sites at the edges of the system, and the current as a function of $t$ 
is measured on the two links connecting the central region to the two outer 
portions of the system. The amplitude of $J(t)$, defined as the average of 
these two currents, scales approximately linearly with the number of sites 
on which the bias potential $\Delta V$ is applied. 
As it is well known,\cite{alhassanieh} $J(t)$ follows an oscillatory 
evolution with time due to the open boundary conditions used 
in the DMRG process (electrons cannot leave the system). Thus, 
to obtain a measure of the conductance, the average of $J(t)/\Delta V$ 
over the first half-period of the oscillation is considered.\cite{alhassanieh}

\section{Results at zero temperature}
\label{T=0}

In this section, the main results obtained in our numerical simulations 
will be described.

\subsection{All-manganite trilayer geometry \\
 using classical $t_{\rm 2g}$ spins}

\subsubsection{Even number of sites in the barrier}\label{even}

\begin{figure}[hbt]
\includegraphics[clip=true,width=8.5cm,angle=-0]{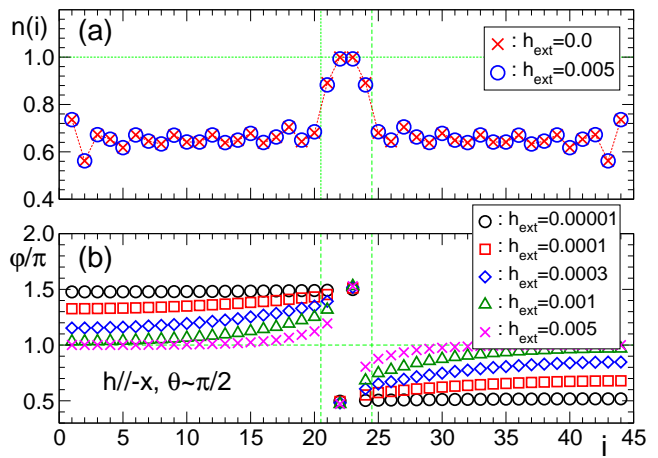}
\begin{center}
\caption{
(Color online) 
(a) Local electron density $n(i)$ and (b) classical spin orientation 
${\bf S}_i=(\theta_i,\varphi_i)$ 
($0\le\theta_i\le\pi$, $0\le\varphi_i\le2\pi$) 
in spherical coordinates vs. site location $i$, on a trilayer 
chain with $L^{(\rm L)}=L^{(\rm R)}=20$ sites in each lead and 
$L^{(\rm C)}=4$ sites in the center. 
The 1D DE model with $J_{\rm H}$=$8.0$ and $\alpha$=$1.0$ is used, and 
the results are obtained via an optimization method at zero temperature. 
Lead's positive background charge is $n_+^{(\rm L)}$=$n_+^{(\rm R)}$=$0.65$, 
and a central positive background charge is $n_+^{(\rm C)}$=$1.0$. 
A magnetic field is applied in the direction opposite to $x$-direction 
({\bf x}), and 
the magnitude ($h_{\rm ext}$) is shown in the figures. 
In (b), $\theta_i\sim\pi/2$ for all $h_{\rm ext}$'s. 
As $h_{\rm ext}$ increases, 
the spins gradually rotate toward $-{\bf x}$ in the $xy$ plane, and finally 
they align ferromagnetically.  The positions of the two interfaces are 
denoted by vertical dashed lines. 
}
\label{den_T0}
\end{center}
\end{figure}

Figure \ref{den_T0} shows the zero-temperature optimization results 
corresponding to a 1D trilayer geometry [Fig.~\ref{spin} (a)] 
consisting of 20 sites in each lead, and 4 sites in the middle. 
The positive charge, regulated by $n_{+}$, is 0.65 at each lead and 
1.0 in the middle (in $e$=1 units). 
The strength of the long-range Coulomb interaction is chosen as $\alpha$=1.0. 
Figure \ref{den_T0}(a) shows the converged local electronic density $n(i)$ 
vs. the site location $i$, along the 44 sites chain. In the leads, the 
electronic density closely matches the expected result 0.65. The 
oscillations at the end of the chain near $i$=1 and 44 are due to 
Friedel oscillations. In the 4-sites center, 2 of the sites have $n(i)$ 
very close to 1, while the other 2 have a smaller density due to the 
charge-transfer effect of the long-range forces. The overall electronic 
density profile is reasonable and in agreement with qualitative expectations.

The lower panel Fig.~\ref{den_T0}(b) contains the most important result 
for this particular geometry. There, the orientations of the classical 
$t_{2g}$ spins are shown via the spherical coordinates angle $\varphi$, 
as a function 
of position $i$. Note first that each lead has classical $t_{\rm 2g}$ spins 
polarized in a ferromagnetic state, namely all sites have approximately the 
same $\varphi$, as expected from the well-known double-exchange 
mechanism.~\cite{zener} 
This ferromagnetism at the leads is compatible with phase diagrams gathered 
in previous studies.\cite{yunoki} However, in the near absence of 
magnetic fields, $h_{\rm ext}$=0.00001, the angle $\varphi$ of the two 
ferromagnetic 
leads differs by $\pi$, signaling an anti-parallel arrangement of magnetic 
moments between 
the two ferromagnetic leads (the tiny field was used simply to orient 
the spins along a direction which simplifies the discussion, and it does 
not have any other important effect). This anti-parallelism is in excellent 
agreement with the expected results based on the introductory discussion: 
if the number of sites in the antiferromagnetic barrier is 
even, then an anti-parallel orientation of ferromagnetic moment between 
the leads should occur. In the barrier region, the classical spins are 
arranged in an antiferromagnetic pattern, since there $n(i)$$\sim$1 and an 
antiferromagnetic spin orientation is preferred.~\cite{yunoki}

The results become even more interesting as the magnetic field is increased. 
In this case, the relative orientation of the ferromagnetic moments in the 
leads changes very rapidly, as shown in Fig.~\ref{den_T0}(b). For fields as 
small as $h_{\rm ext}$=0.005 (in units of the hopping $t$), the magnetic 
moments 
of the ferromagnetic leads are already nearly aligned. As shown below, 
this produces drastic changes in the conductance of the ensemble. Results 
for intermediate values of the magnetic field, also in Fig.~\ref{den_T0}(b), 
show that the transition from anti-parallel to parallel orientation of the 
ferromagnetic lead's moments is smooth and, moreover, noticeable changes 
can be observed even for fields as tiny as $h_{\rm ext}$=0.0001 
(a discussion of how small this field is in physical units is below).

\begin{figure}[hbt]
\includegraphics[clip=true,width=8.5cm,angle=-0]{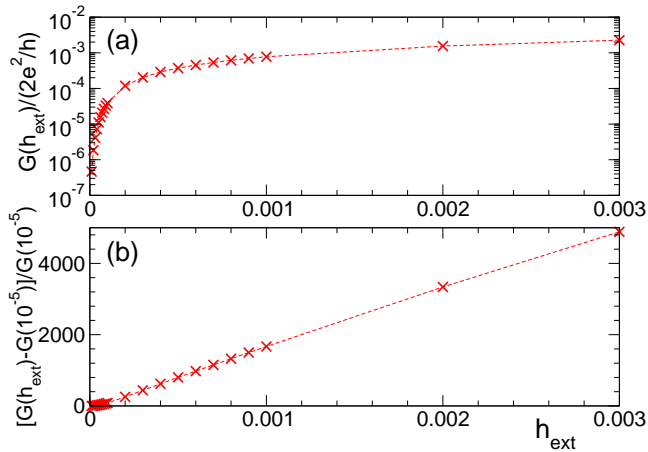}
\begin{center}
\caption{
(Color online) 
(a) Conductance $G(h_{\rm ext})$ in units of $2e^2/h$ ($h$: Planck constant) 
and 
(b) magnetoresistance (MR) ratio, as a function of an applied magnetic 
field $h_{\rm ext}$. 
The results are calculated for the 1D DE model with $J_{\rm H}$=$8.0$, 
$\alpha$=$1.0$, and $T$=$0.0$. Positive background charges are 
$n_+^{(\rm L)}$=$n_+^{(\rm R)}$=$0.65$ and $n_+^{(\rm C)}$=$1.0$, and 
leads and central region sizes are $L^{(\rm L)}$=$L^{(\rm R)}$=$20$ 
and $L^{(\rm C)}$=$4$, respectively. 
To estimate the MR ratio, defined in the vertical axis of (b), 
$G(h_{\rm ext})$ with $h_{\rm ext}=10^{-5}$, instead of $G(h_{\rm ext}=0)$, is 
used since the conductance at $h_{\rm ext}$=0 vanishes. 
Note that with increasing $h_{\rm ext}$ the conductance $G(h_{\rm ext})$, 
and the MR ratio, 
increases almost linearly except for very small $h_{\rm ext}$ ($\alt 0.0003$). 
}
\label{cond_T0}
\end{center}
\end{figure}

The relative orientation of the magnetic moments of the leads, and their 
rotation with 
magnetic fields, induce substantial modifications in the conductance and a 
concomitant large magnetoresistance. Figure~\ref{cond_T0}(a) shows the 
conductance of the trilayer vs. magnetic field $h_{\rm ext}$. At fields 
zero or very 
small, the anti-parallel orientation of the lead's moments produces a very 
small conductance. This is reasonable since the spin orientation which 
can conduct in one lead, is blocked by the anti-parallel lead. However, 
as $h_{\rm ext}$ increases, there are substantial changes in the conductance. 
For fields as small as $h_{\rm ext}$=0.0002, the conductance has changed by 
about 3 orders of magnitude already, at least for the particular system 
studied here. Note that if $t$ is assumed to be 0.1~eV, 
then $h_{\rm ext}$=0.0001 is approximately 0.1~T. The conductance increases 
further, 
by more than an order of magnitude, by  increasing $h_{\rm ext}$ toward 
the value 
0.003 where the moments of the leads become essentially parallel. 
The conductances 
remain in absolute value much smaller than the perfect conductance 
$2e^2/h$ ($h$: Planck constant), but its relative changes can be 
large, as illustrated in Fig.~\ref{cond_T0}(b). In the widely used 
definition for conductance changes, which has the zero field conductance 
in the denominator, the magnetoresistance ratio can be very large, 
and it reaches almost 200,000\% at $h_{\rm ext}$=0.001.

To give a better perspective of the importance of these numbers, note that 
in a recent numerical study\cite{sen} of the same model used for the 
trilayer geometry but defined on a finite two-dimensional cluster without 
interfaces, the CMR phenomenon was observed for larger magnetic fields 
$h_{\rm ext}$=0.05. In this case the MR magnitude was 10,000\% at the best. 
Thus, the trilayer geometry investigated here certainly produces 
a more dramatic effect at smaller fields than in the bulk simulations.

\begin{figure}[hbt]
\includegraphics[clip=true,width=8.5cm,angle=-0]{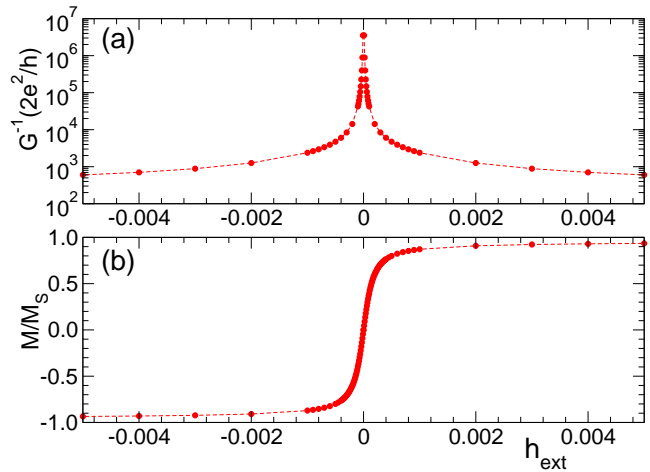}
\caption{
(Color online) 
(a) Resistance (inverse of conductance) vs. magnetic field $h_{\rm ext}$ 
for an even 
number of sites $L^{(\rm C)}$=4 in the central barrier at zero 
temperature. The leads each has $L^{(\rm L)}$=$L^{(\rm R)}$=20 sites, and 
the positive background charges are $n_+^{(\rm L)}$=$n_+^{(\rm R)}$=$0.65$ 
and $n_+^{(\rm C)}$=$1.0$ in the leads and central region, respectively. 
The couplings used are $J_{\rm H}$=$8.0$, and $\alpha$=$1.0$. Notice that 
the large changes in resistance occur at very small magnetic fields. 
(b) Total magnetization $M$ of the classical $t_{2g}$ spins vs. $h_{\rm ext}$, 
indicating the rapid development with $h_{\rm ext}$ of a net magnetization. 
$M_S$ is the maximum possible magnetization.
}
\label{R-even}
\end{figure}

To help experimental readers who are more used to resistance plots, 
in Fig.~\ref{R-even}(a), the resistance $R$ of the trilayer with an even 
number of sites in the barrier is shown as a function of magnetic field 
$h_{\rm ext}$. 
Notice the rapid change in $R$ at small $h_{\rm ext}$, which is the main 
result of 
this paper. 
In experimental similar plots, such as those reported for STO and LAO as 
barriers, hysteresis loops are often observed in TMR vs. $h_{\rm ext}$, 
and moreover 
the effects at tiny fields are negligible.~\cite{bowen,yamada,ishii} 
These differences are caused 
by the anisotropies present in real experiments due to strain in the
samples, as briefly explained in Sec.~\ref{main_appr}, effect that has 
been mainly neglected in our calculations (with the exception 
of the results shown in Fig.~\ref{TMR_s} below). In Fig.~\ref{R-even}(b), 
magnetization $M$ of the classical $t_{2g}$ spins for the trilayer 
ensemble vs. $h_{\rm ext}$ is shown. 
Compatible with the spin arrangements already described, 
there is no net magnetization $M$ at $h_{\rm ext}$=0, 
since the contributions of 
the ferromagnetic leads cancel out. However, with increasing 
$h_{\rm ext}$, there is a rapid increase in $M$ because the lead's 
magnetic moments are aligning in the same direction. 
The subsequent small increases in $M$ and decrease in $R$, say 
for $h_{\rm ext}$=0.001 or larger, are due to further fine alignment 
of the spins 
close to the central region (the spin orientation in the central region 
remains antiferromagnetic, which eventually turns to FM with much larger 
$h_{\rm ext}$). Thus, the effective height of the barrier 
melts as $h_{\rm ext}$ (or $T$, as discussed in the next section) increases 
in this system.

\subsubsection{Odd number of sites in the barrier}

\begin{figure}[hbt]
\includegraphics[clip=true,width=8.5cm,angle=-0]{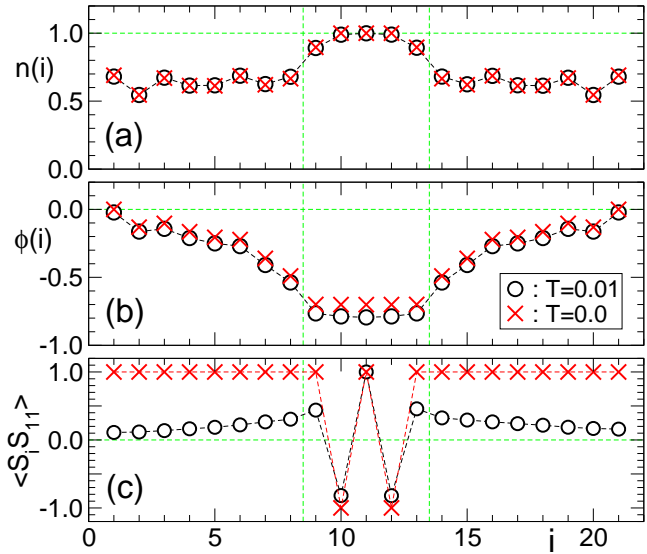}
\begin{center}
\caption{
(Color online) 
(a) Local electron density $n(i)$, (b) electrostatic potential $\phi(i)$, 
and (c) classical $t_{2g}$ spin correlation function 
$\langle {\bf S}_i\cdot{\bf S}_j \rangle$ from the central site 
at $j$=$11$. 
The Monte Carlo techniques and optimization method are used for the 1D DE 
model with $J_{\rm H}$=$8.0$ and $\alpha$=$2.0$ at $T$=$0.01$ and $T=0$, 
respectively. The positive background charges are 
$n_+^{(\rm L)}$=$n_+^{(\rm R)}$=$0.625$ and $n_+^{(\rm C)}$=$1.0$, and the 
system size used is $L^{(\rm L)}$=$L^{(\rm R)}$=$8$ and 
$L^{(\rm C)}$=$5$. 
Here, $\phi(i)$ is self-consistently determined. 
The interface positions are denoted by vertical dashed lines. 
 }
\label{den}
\end{center}
\end{figure}

The numerical results obtained for the case of an odd number of sites in 
the barrier are very different from those reported thus far. 
In Fig.~\ref{den}, the results for the case of $L^{(\rm C)}=5$ sites 
in the barrier are reported. Fig.~\ref{den} (a) shows the electronic density 
$n(i)$ vs. $i$, which is similar to the results for the case of 
$L^{(\rm C)}=4$ sites in the barrier [Fig.~\ref{den_T0} (a)]. 
The electrostatic potential for electrons $\phi(i)$, 
caused by the long-range Coulomb 
interactions, shown in Fig.~\ref{den} (b), is also canonical: in the 
central region more electrons are expected to accumulate since there 
$n_{+}(i)$ is larger than in the leads. 
The important qualitative difference with the previous results in 
Sec.~\ref{even} is presented in Fig.~\ref{den} (c). 
Here, the spin correlation functions for the classical $t_{2g}$ spins 
indicate that the magnetic moments of the leads 
are parallel to each other, even in the absence of magnetic fields. 
This is in agreement with the qualitative scenario for these geometries 
described in the introduction (Sec.~\ref{idea}). 
As a consequence, the case of an odd number of sites in the barrier does 
$not$ present the same huge magnetoresistance at small magnetic fields 
as for the even case.

\begin{figure}[hbt]
\includegraphics[clip=true,width=8.5cm,angle=-0]{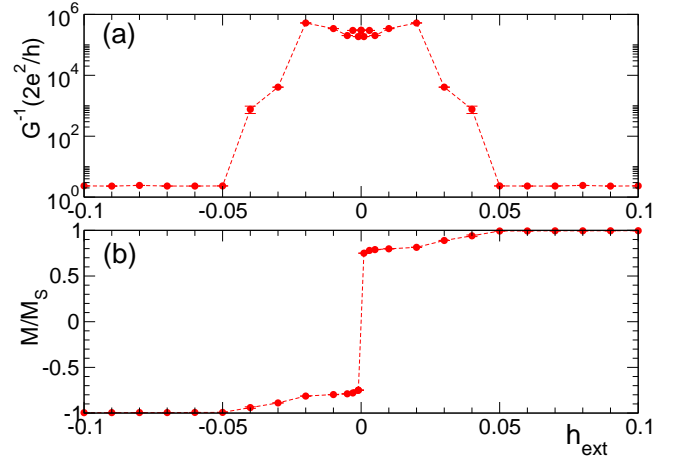}
\caption{
(Color online) 
(a) Resistance (inverse of conductance) vs. magnetic 
field $h_{\rm ext}$ for an odd 
number of sites $L^{(\rm C)}$=5 in the central barrier at $T$=1/800. 
The leads have $L^{(\rm L)}$=$L^{(\rm R)}=8$ sites, and 
the positive background charges are $n_+^{(\rm L)}$=$n_+^{(\rm R)}$=$0.625$ 
and $n_+^{(\rm C)}$=$1.0$ for the leads and central region, respectively. 
The couplings used are $J_{\rm H}$=$8.0$ and $\alpha$=$1.0$. 
(b) Total magnetization $M$ of the classical $t_{2g}$ spins vs. magnetic 
field $h_{\rm ext}$, indicating that the system is overall ferromagnetic 
at all 
magnetic fields. $M_S$ is the maximum possible magnetization.
$\phi(i)$ is determined self-consistently at $T$=$0.2$ and 
$h_{\rm ext}=0$, and then used for other temperatures and $h_{\rm ext}$ 
(see more details in the text). 
}
\label{R-odd}
\end{figure}

The resistance $R$ vs. magnetic field $h_{\rm ext}$ is shown in 
Fig.~\ref{R-odd}(a). 
As in the case of the even $L^{(\rm C)}$, clearly $R$ decreases with 
increasing $h_{\rm ext}$. However, the scales involved are very different. 
While 
for $L^{(\rm C)}$ even, there are huge changes at $h_{\rm ext}$ as small as 
0.0001 (Fig.~\ref{R-even}), for the $L^{(\rm C)}$ odd case the resistance 
remains almost the same up to $h_{\rm ext}$=0.02. The reason is that the 
magnetic field does not need to align the magnetic moments of the leads 
in this case 
(they are already aligned), and moreover the up-down-up arrangement of 
the central region is compatible with the parallel leads' moments. 
The only modification needed in the spin arrangement is the correction 
in the orientation of the central spins that are pointing the wrong way. 
This process takes place between fields $h_{\rm ext}$=0.02 and 0.05 
approximately. 
After that, the entire system is ferromagnetic and the resistance remains 
constant. For completeness, Fig.~\ref{R-odd}(b) shows the total 
magnetization of the classical $t_{2g}$ spins. As expected, initially, $M$ 
is very robust, and it becomes saturated at $h_{\rm ext}\sim0.05$.

\subsection{All-manganite trilayer geometry \\
using quantum $t_{\rm 2g}$ spins}

The existence of parallel or anti-parallel arrangements of the 
magnetic moments in the left and right leads is also investigated for 
the 1D DE model with 
quantum localized $t_{2g}$ spins ${\bf S}_i$ [Eq.~(\ref{model})] at $T=0$ 
by the DMRG technique. For numerical simplicity, spin-1/2 is 
employed, instead of the realistic value 3/2. 

\begin{figure}[hbt]
\includegraphics[clip=true,width=7cm]{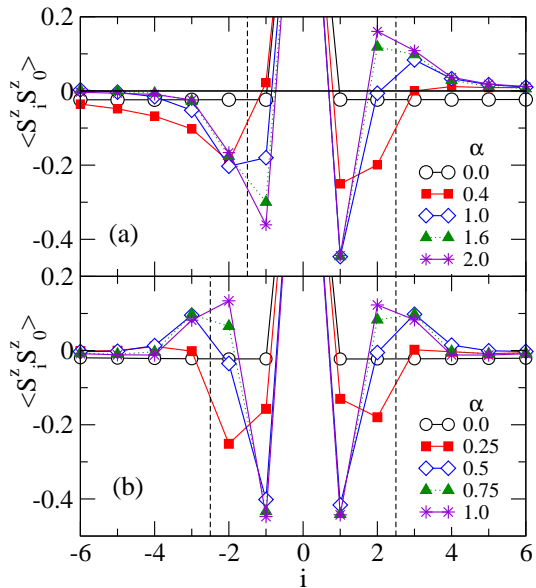}
\begin{center}
\caption{(Color online)
Spin correlation functions $\langle S^z_i S^z_j\rangle$ between 
quantum $t_{2g}$ spins (spin-1/2), calculated from the center of the 
chain at site $j=0$, for the 
1D DE model with $J_{\rm H}=32$ and various $\alpha$'s (indicated in the 
figure) at $T=0$. 
(a): $L^{(\rm L)}=L^{(\rm R)}=12$ and $L^{(\rm C)}=4$, and the positive 
background charge densities are $n_+^{(\rm L)}$=$n_+^{(\rm R)}$=0.5 and 
$n_+^{(\rm C)}$=1.0. (b): $L^{(\rm L)}=L^{(\rm R)}=14$ and $L^{(\rm C)}=5$, 
and the positive background charge densities are 
$n_+^{(\rm L)}$=$n_+^{(\rm R)}$=0.571 and $n_+^{(\rm C)}$=1.0.
The normalization $\langle S^z_0 S^z_0 \rangle$=$1$ is adopted.
}
\label{jose.fig2}
\end{center}
\end{figure}

In Fig.~\ref{jose.fig2}, the ($z$-component) spin correlation functions 
$\langle S^z_i S^z_j\rangle$ are reported for a typical set of parameters, 
choosing one of the central-region spins ($j=0$) as a reference. 
There are 4 and 5 sites in the central region 
in Fig.~\ref{jose.fig2} (a) and (b), respectively. 
In the case of even number of central sites 
[Fig.~\ref{jose.fig2} (a)], at least for large values of $\alpha$ such as 2, 
the antiferromagnetic spin correlations in the central region are clearly 
observed, and moreover the magnetic moments of the left and right leads tend 
to align 
anti-parallel. However, quantum fluctuations, enhanced 
by the spin-1/2 and one-dimensionality nature of the model, make the 
spin correlations at longer distances weaker, and the correlations become 
very small for $|i|\agt4$. Similarly, quantum fluctuations reduce spin 
correlations at large distances also for odd number of central sites, 
shown in Fig.~\ref{jose.fig2} (b). However, in this case, a tendency of 
parallel alignment of the magnetic moments between the left and right 
leads can still be observed. 
Thus, overall features observed in the case of quantum $t_{2g}$ spins are 
in good qualitative agreement with the ones observed in the classical 
$t_{2g}$ spin case. Realistic two- or three-dimensional manganite 
trilayers are expected to behave more similarly to the classical $t_{2g}$ 
spin case than the quantum spin-1/2 one-dimensional case, where the quantum 
fluctuations are the strongest. In fact, quantum fluctuations appear 
detrimental for the performance of the device proposed in this paper. 
Higher dimensional arrangements will have a stronger tendency to spin 
order, as the classical spins do here.

\begin{figure}[hbt]
\includegraphics[clip=true,width=7cm]{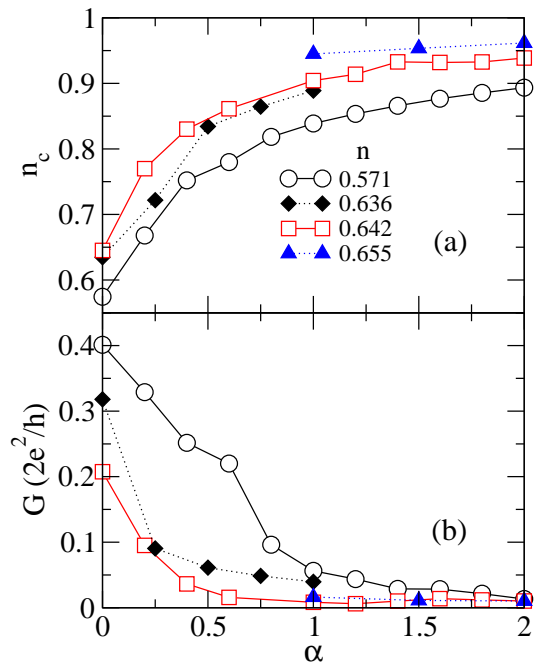}
\begin{center}
\caption{(Color online) 
(a) Electron density in the central section $n_c$ 
($=\sum_{i\subset{\rm C}}n_i/L^{(\rm C)}$), 
and (b) conductance $G$, as a function of $\alpha$, for various values of the
total electron density $n$ indicated in the plots. Results are for the 
1D DE model with quantum $t_{2g}$ spin (spin-1/2) and 
$J_{\rm H}=32$ at 
$T=0$, calculated using the DMRG method. The other parameters are
$L^{(\rm L)}=L^{(\rm R)}=12$, $L^{(\rm C)}=4$ (circles and squares),  
$L^{(\rm L)}=L^{(\rm R)}=14$, $L^{(\rm C)}=5$ (diamonds), 
and 
$L^{(\rm L)}=L^{(\rm R)}=12$, $L^{(\rm C)}=5$ (triangles). 
In all cases, the positive background charges are $n_+^{(\rm C)}$=1.0 and 
$n_+^{(\rm L)}$=$n_+^{(\rm R)}$. 
}
\label{jose.fig1}
\end{center}
\end{figure}

With decreasing $\alpha$, the electron density in the central region is 
substantially 
reduced [see Fig.~\ref{jose.fig1} (a)], and eventually the antiferromagnetic 
correlations at short distances within the central region disappear for 
both even and odd number of
sites in the center (Fig.~\ref{jose.fig2}). Together with this effect, an 
increase in the conductance with decreasing $\alpha$ is observed as 
shown in Fig.~\ref{jose.fig1} (b). 
Here the conductance is calculated with the time-dependent DMRG 
technique~\cite{tDMRG,alhassanieh}, already explained in Sec.~\ref{dmrg2}.  
From these results, in order for the central region to play the role of 
a tunneling barrier with low conductance, it is clear that 
the spins in the central region must be antiferromagnetically aligned, 
and the electronic density there must be close to 1, which is achieved 
only by a large $\alpha$, i.e., strong Coulomb interactions.

\begin{figure}
\includegraphics[clip=true,width=8cm]{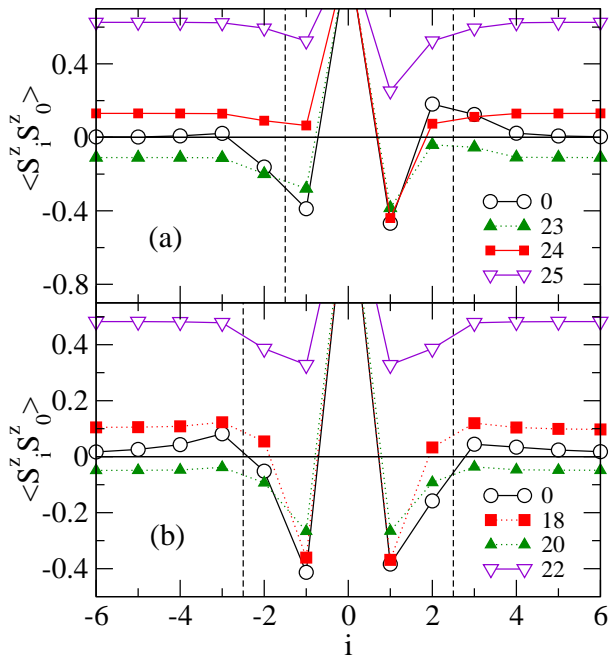}
\begin{center}
\caption{(Color online)
Spin correlation functions $\langle S^z_i S^z_j\rangle$ between 
quantum $t_{2g}$ spins (spin-1/2), calculated from the center of the 
chain at site 
$j=0$, for the 1D DE model with $J_{\rm H}=32$, $\alpha$=1.0, and various 
total $S^z$ (i.e., total magnetization) indicated in the figure. 
The calculations are performed with the DMRG method at $T=0$. 
(a): $L^{(\rm L)}=L^{(\rm R)}=12$, $L^{(\rm C)}=4$, and total electron 
density $n$=$0.571$. (b): $L^{(\rm L)}=L^{(\rm R)}=12$, $L^{(\rm C)}=5$, 
and $n$=$0.586$. The positive background charge densities 
are $n_+^{(\rm L)}$=$n_+^{(\rm R)}=0.5$ and $n_+^{(\rm C)}$=1.0. 
The normalization $\langle S^z_0 S^z_0 \rangle$=$1$ is adopted. 
}
\label{jose.fig3}
\end{center}
\end{figure}

The influence of magnetic fields is also studied in 
Fig.~\ref{jose.fig3}. Here, instead of applying an external magnetic field, 
the total $z$-component of the spins (i.e., total magnetization), which is 
a good quantum number of the system, is varied as a parameter. 
Fig.~\ref{jose.fig3} (a) indicates that the anti-parallel correlations 
of the lead's spins, for the case of even number $L^{(\rm C)}=4$ of central 
sites, becomes parallel with increasing total magnetization. 
This is in qualitative agreement with the observation in the previous 
subsection, indicating that classical and quantum $t_{\rm 2g}$ 
spins behave qualitatively similarly in this system. The same occurs 
for the case of an odd number of central sites, shown in 
Fig.~\ref{jose.fig3} (b). There is a small caveat to mention here: 
(i) the behavior with increasing total magnetization is not monotonous, 
and (ii) some of the spin correlations in Fig.~\ref{jose.fig3} 
(as well as in Fig.~\ref{jose.fig2}) 
are small, which might be explained by canting 
effects or by nearly orthogonal spin configurations. These issues are not 
further explored here, since they do not appear in the classical 
$t_{2g}$ spin 
simulations which seem more realistic to describe manganites. 
In spite of these caveats, it is clear that qualitatively the similar 
feature is observed for both models with classical and with quantum 
$t_{2g}$ spins, the key feature of having anti-parallel magnetic lead 
configurations for an even number of central sites.

%


\section{Influence of temperature}
\label{Tnonzero}

The large MR effect at low temperature observed in the trilayer geometry 
described in this paper is an interesting effect worthy of experimental 
confirmation. However, for practical applications, this large MR effect 
should survive up to high temperatures, more specifically room temperature 
or above. In fact, previous experimental realizations of TMR devices 
suffered a rapid degradation with increasing temperatures,~\cite{bowen} 
as mentioned in the introduction (Sec.~\ref{intro_stj}). 
Unfortunately, our proposed trilayer system also has problems in this 
respect, as shown below, but possible avenues to solve this issue are 
discussed.

\begin{figure}[hbt]
\includegraphics[clip=true,width=8.5cm,angle=-0]{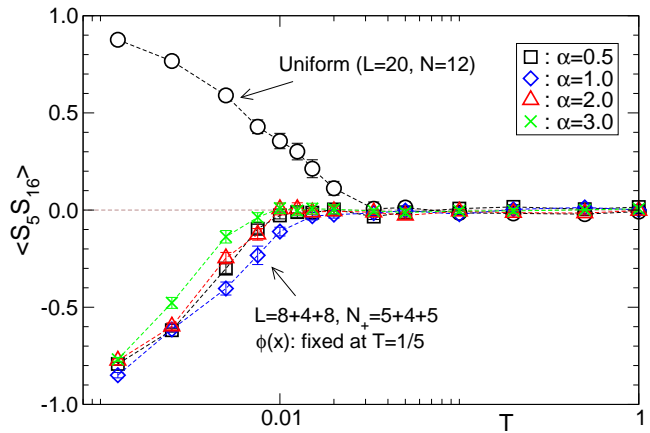}
\begin{center}
\caption{
(Color online) 
Classical $t_{2g}$ spin correlation functions 
$\langle {\bf S}_i\cdot{\bf S}_j \rangle$ between site 
$i$=$5$ (belonging to the left lead) and site $j$=$16$ 
(belonging to the right lead), as a function of temperature $T$. 
Monte Carlo simulations are used for the 1D DE model with 
$J_{\rm H}$=$8.0$, $n_+^{(\rm L)}$=$n_+^{(\rm R)}$=$0.625$, 
$n_+^{(\rm C)}$=$1.0$, $L^{(\rm L)}$=$L^{(\rm R)}$=$8$, and $L^{(\rm C)}$=$4$. 
The values of $\alpha$ used are indicated in the figure. 
Here, $\phi(i)$ is determined self-consistently at $T$=$0.2$ and is 
subsequently used for other lower temperatures. For comparison, 
the classical $t_{2g}$ spin correlation functions for a uniform DE model 
with $J_{\rm H}$=$8.0$, $L$=$20$, and the number of electrons $N$=$12$ are 
also plotted by circles. 
}
\label{spin_alpha}
\end{center}
\end{figure}

In Fig.~\ref{spin_alpha}, classical $t_{2g}$ spin correlation functions for 
two particular sites are shown as a function of temperature $T$, computed 
with the Monte Carlo technique. One of the chosen sites is in the left 
lead, site 5, and the other is in the right lead, site 16, and they are 
symmetrically arranged with respect to the center for the $L=8+4+8$ cluster 
used. For comparison, the same correlation is shown for the case when the 
barrier is removed and the entire system is now a unique ferromagnetic 
metal.~\cite{yunoki} The latter decays with temperature as shown in 
Fig.~\ref{spin_alpha}, indicating that the long-distance ferromagnetic 
tendencies survive up to $T_{\rm C}\sim$0.03 approximately. Although this 
should not be considered as a critical temperature, due to the one 
dimensionality of the problem which introduces strong fluctuations at 
finite temperatures, at least it provides a good indicator of 
the strength of ferromagnetism as $T$ is increased. Weak couplings 
into higher dimensional structures will likely stabilize this characteristic 
temperature into a true critical 
temperature.~\cite{dagotto-CMR,dagotto-book,yunoki}

The same spin correlation functions but now in the presence of 
the barrier is also shown in Fig.~\ref{spin_alpha} for various values of 
$\alpha$. At low temperatures, these correlation functions have the opposite 
sign (minus) as compared to the ones without the barrier. This is simply 
because in this trilayer case there is an 
even number of sites in the central region, and therefore the magnetic 
moments between the leads align antiferromagnetically (Sec.~\ref{T=0}). 
Here the calculations are carried out by first obtaining the electrostatic 
potential $\phi(i)$ at high temperature, where the spins are not ordered, 
and then keeping this potential the same as the temperature is reduced. 
This procedure considerably alleviates the numerical effort, particularly 
regarding the Poisson's equation iterations, and tests in small systems have 
shown that this trick does not alter the results qualitatively. 
As also seen in Fig.~\ref{spin_alpha}, the spin correlation functions 
obtained by this method have a relatively minor dependence with $\alpha$. 
The main point to remark is that these spin correlations become negligible 
at a temperature $T^*\sim0.01$, which is considerably smaller than 
the relevant temperature of the pure ferromagnetic system without the 
barrier. As a consequence, it is clear that a strong similarity with 
previous experimental results for STO barriers~\cite{bowen} may exist in 
this case. Expressed qualitatively, our results indicate that there is a 
temperature scale $T^*$, much smaller than the critical temperature 
$T_{\rm C}$, where the orientation of the leads' magnetic moments ceases to 
be antiferromagnetic. As shown below, this is correlated with important 
changes in the magnetoresistance effect. In fact, the large MR effect at 
low temperature seems to occur due to the existence of an effective 
coupling $J_{\rm eff}$ which produces the anti-parallel alignment of the 
leads' magnetic moments. This is reasonable, since $J_{\rm eff}$ is an 
effective weak coupling across a tunneling insulating barrier. When the 
temperature is of the order of this coupling or larger, the anti-parallel 
arrangement is no longer preferable and in our model the leads' moments 
rotate freely with respect to one another.

\begin{figure}[hbt]
\includegraphics[clip=true,width=8.5cm,angle=-0]{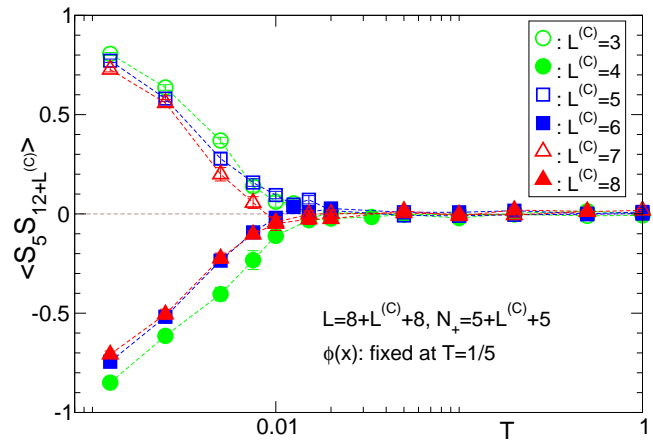}
\begin{center}
\caption{
(Color online) 
Classical $t_{2g}$ spin correlation functions 
$\langle {\bf S}_i\cdot{\bf S}_j \rangle$ between sites 
$i$=$5$ (belonging to the left lead) and site 
$j$=$12+L^{(\rm C)}$ (belonging to the right lead), for the 1D DE model with 
$J_{\rm H}$=$8.0$, $\alpha$=$1.0$, $n_+^{(\rm L)}$=$n_+^{(\rm R)}$=$0.625$, 
$n_+^{(\rm C)}$=$1.0$, $L^{(\rm L)}$=$L^{(\rm R)}$=$8$, and 
$L^{(\rm C)}$=$3,4,\cdots,8$. 
Here, $\phi(i)$ is determined self-consistently at $T$=$0.2$ and then 
used for other temperatures.}
\label{spin_C}
\end{center}
\end{figure}

A systematic study of the influence of the size of the central region 
$L^{(\rm C)}$ on the classical $t_{2g}$ spin correlations between the leads 
is reported in Figure~\ref{spin_C}. As expected (Sec.~\ref{idea}), for 
$L^{(\rm C)}$ even (odd) these spin correlations are negative (positive) at 
low temperatures, indicating strong antiferromagnetic (ferromagnetic) 
correlations between the ferromagnetic leads' moments. The magnitude
of this correlation decreases as $L^{(\rm C)}$ increases, in agreement with the
expected reduction of the effective coupling $J_{\rm eff}$ discussed above. 
Thus, the thinner the barrier is, the better 
the temperature effects become in the trilayer geometry proposed here, 
namely the higher $T^*$ is.

\begin{figure}[hbt]
\includegraphics[clip=true,width=8.5cm,angle=-0]{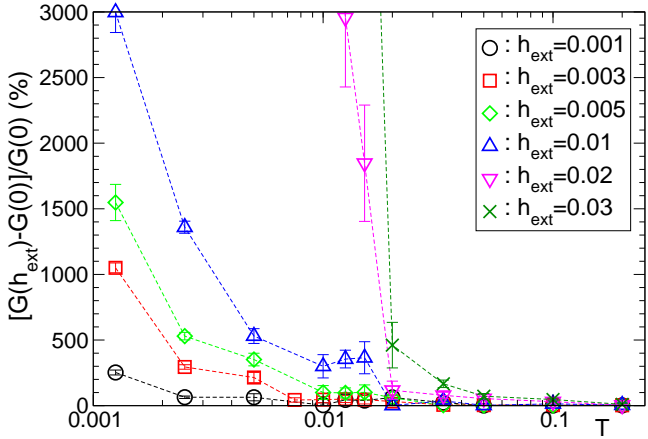}
\begin{center}
\caption{
(Color online) 
Temperature dependence of the 
MR ratio $[(G(h_{\rm ext})-G(0))/G(0)]\times100$ for the 1D DE model with 
$J_{\rm H}$=$8.0$, $\alpha$=$1.0$, $n_+^{(\rm L)}$=$n_+^{(\rm R)}$=$0.625$, 
$n_+^{(\rm C)}$=$1.0$, $L^{(\rm L)}$=$L^{(\rm R)}$=$8$, and 
$L^{(\rm C)}$=$4$. Applied magnetic fields $h_{\rm ext}$ are indicated 
in the figure. 
$\phi(i)$ is determined self-consistently at $T$=$0.2$ and used for the other 
temperatures. 
Note that the effect reported at $T$=0 in Fig.~\ref{cond_T0} is 
much larger in magnitude of the order of 200,000\% at $h_{\rm ext}$=0.002 
(in that figure the factor 100 was not used as in here). In the range of 
temperatures shown in this figure, the original $T$=0.0 nearly 
perfect antiferromagnetic alignment of the leads' magnetic moments is 
already lost at $h_{\rm ext}$=0, and thus the changes in resistances are not 
as dramatic as observed at $T=0$.
}
\label{TMR}
\end{center}
\end{figure}

Our discussion thus far has been based on the spin correlations at finite 
temperatures. It was concluded that the large resistance caused by the 
anti-parallel configuration of the ferromagnetic leads' moments does not 
survive all the way to the ferromagnetic critical temperatures of the leads. 
This is indeed observed in Fig.~\ref{TMR}, where at $T^*$ the MR effect is 
reduced to 
zero at fields such as $h_{\rm ext}$=0.001-0.003, while at zero temperature 
[Fig.~\ref{cond_T0} (b)] the MR effect was huge at the same fields. 
However, even though the MR effect is truly enormous at very low temperature, 
at higher temperatures it is not negligible, at least in the several Teslas 
scale of magnetic fields $h_{\rm ext}$. For example, in Fig.~\ref{TMR}, 
the MR ratio 
is plotted vs. temperature, for different values of $h_{\rm ext}$. 
Note that  $[G(h_{\rm ext})-G(0)]/G(0)$ is the same as the more standard 
definition $[R(0)-R(h_{\rm ext})]/R(h_{\rm ext})$, where the resistance 
$R(h_{\rm ext})$=$1/G(h_{\rm ext})$. 
For fields such as $h_{\rm ext}$=0.02 and 
0.03 -- corresponding to 20 and 30~T, 
respectively, if it is assumed $t$$\sim$1,000~T -- the MR ratio can be 
as large as 100 or 500\% at temperatures of the order of $T_{\rm C}$/2.

\begin{figure}[hbt]
\includegraphics[clip=true,width=7.5cm,angle=-0]{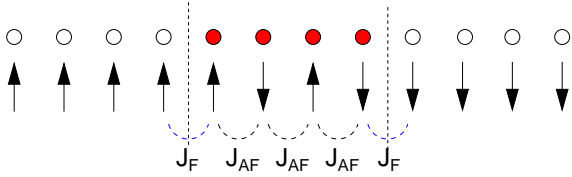}
\begin{center}
\caption{
(Color online) 
Schematic representation of the 1D DE model [Eq.~(\ref{model})] for 
a LSMO/LMO/LSMO trilayer (see also Fig.~\ref{spin}) with additional 
Heisenberg couplings between classical $t_{2g}$ spins, which are introduced 
to simulate anisotropies. 
Antiferromagnetic $J_{\rm AF}$, as well as ferromagnetic $J_{\rm F}$, 
couplings are added only in the central region and the bonds connecting to 
the central region, as indicated in 
the figure. The two interface positions are denoted by vertical dashed 
lines. 
}
\label{spin_2}
\end{center}
\end{figure}

In the calculations described so far, no ``easy axis'' has been chosen. 
In other words, no anisotropies were introduced. However, the introductory 
discussion suggests that many materials, particularly when in thin-film 
form, do have an easy axis mainly due to the influence of the substrate 
(Sec.~\ref{main_appr}). If due to this effect the magnetic moments of the 
leads cannot rotate ``freely'' with respect to one another as isotropic 
Heisenberg vectors, but are pointing along a particular direction as 
Ising variables, the temperature scale $T^*$ above which the magnetic 
moments of the leads become uncorrelated should increase: Heisenberg-like 
isotropic spins can be disordered by thermal fluctuations more easily 
than Ising-like spins. 
On the other hand, making more ``rigid'' the anti-parallel magnetic 
connection between the leads moments will also prevent their alignment in 
very small magnetic fields: larger magnetic fields may be needed to achieve 
the same effect as before, i.e., large MR effect. 
These competing tendencies will be discussed in more detail below.

\begin{figure}[hbt]
\includegraphics[clip=true,width=8.5cm,angle=-0]{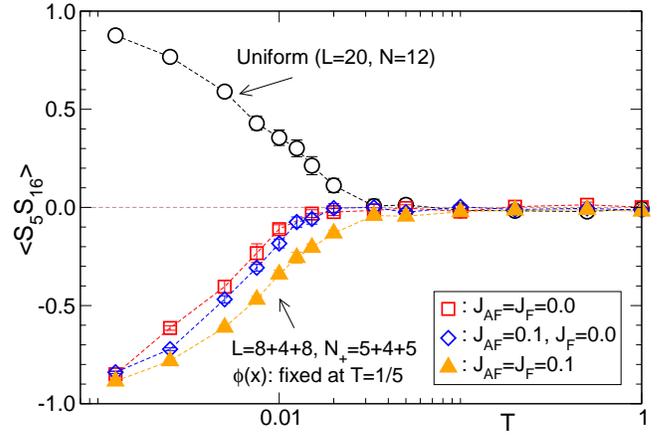}
\begin{center}
\caption{
(Color online) 
Classical $t_{2g}$ spin correlation functions 
$\langle {\bf S}_i\cdot{\bf S}_j \rangle$ between site 
$i=5$ (belonging to the left lead) and site $j=16$ 
(belonging to the right lead), as a function of temperature $T$. 
Monte Carlo simulations are used for the 1D DE model with 
$J_{\rm H}$=$8.0$ and $\alpha=1.0$. The positive background charge 
densities are $n_+^{(\rm L)}$=$n_+^{(\rm R)}=0.625$ and 
$n_+^{(\rm C)}=1.0$, and the cluster used is $L^{(\rm L)}=L^{(\rm R)}=8$ and 
$L^{(\rm C)}=4$. The results with $J_{\rm AF}=J_{\rm F}=0.1$ are shown by 
triangles, and the results with $J_{\rm AF}=0.1$ but no $J_{\rm F}$ are 
denoted by diamonds. 
For comparison, the results for a uniform DE model without the barrier 
(circles) and the results with the central barrier but no 
$J_{\rm AF}$ and $J_{\rm F}$ (squares) are also plotted. These are 
reproduced from Fig.~\ref{spin_alpha}. 
Note that with nonzero $J_{\rm F}$ and $J_{\rm AF}$, the spin correlations 
now vanish at almost the same temperature as in the case without the barrier 
(circles). 
Here, $\phi(i)$ is determined self-consistently at $T$=$0.2$ and is 
subsequently used for other lower temperatures. 
}
\label{spin-special}
\end{center}
\end{figure}

\begin{figure}[hbt]
\includegraphics[clip=true,width=8.5cm,angle=-0]{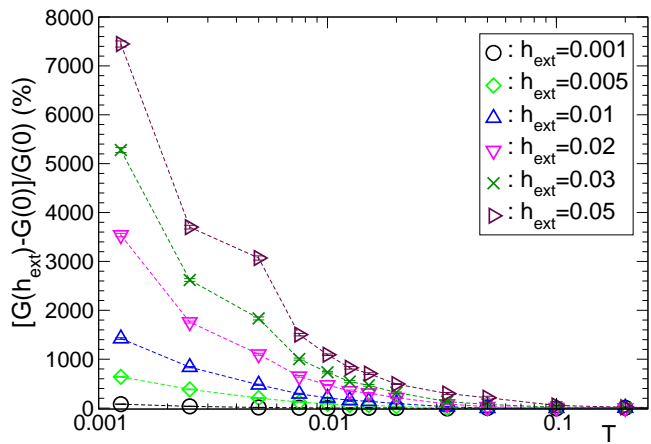}
\begin{center}
\caption{
(Color online) 
The same as in Fig.~\ref{TMR} except that additional couplings 
between the classical $t_{2g}$ spins, $J_{\rm AF}$=$J_{\rm F}$=$0.1$, are 
added to simulate the presence of anisotropies (see Fig.~\ref{spin_2}). 
The applied magnetic fields $h_{\rm ext}$ are indicated in the figure. 
The electrostatic potential $\phi(i)$ is determined self-consistently at 
$T$=$0.2$ and then used for other lower temperatures. }
\label{TMR_s}
\end{center}
\end{figure}

To investigate this issue, extra couplings between classical $t_{2g}$ spins 
are added to the model, as shown in Fig.~\ref{spin_2}. 
Via direct Heisenberg couplings between the classical spins, 
$J_{\rm AF}$ (antiferromagnetic) and $J_{\rm F}$ (ferromagnetic), 
both the antiferromagnetic spin arrangement 
in the central region and the coupling between the center and the leads 
can be made more rigid. That this procedure helps regarding the $T^*$ 
problem is clear in Fig.~\ref{spin-special} where the classical 
$t_{2g}$-spin correlation functions are shown. 
Contrary to the case of weakly coupled leads, now a robust spin correlation 
between them survives up to $T_{\rm C}$. However, the MR effect is not 
improving at very small fields (see Fig.~\ref{TMR_s}). On the contrary, 
larger values of $h_{\rm ext}$ are needed to achieve the same MR effect 
as before 
(for instance, the MR ratio at $h_{\rm ext}$=0.01 without anisotropies 
is 3000\% at 
$T$$\sim$0.001, while it is 1500\% here with anisotropy). 
Thus, as already mentioned briefly, the overall conclusion is that there 
are two 
competing tendencies to consider in these investigations: (i) on one hand, 
a weak coupling among the leads, mediated by the central region, is needed 
in order for a tiny magnetic field to cause a huge effect in transport; 
(ii) on the other hand, for the same reason, a small temperature can 
entirely wash out the effect. Adding anisotropies increases the effective 
coupling between the leads' magnetic moments, thus helping with the 
temperature issue (ii), 
but reciprocally, larger magnetic fields are needed to achieve the same 
MR effect. Investigating the subtle balance between these competing 
tendencies is a great challenge, both for theorists and experimentalists.

\section{Ti-oxide barrier}
\label{STO}

Although the focus of our effort has been on the LSMO/LMO/LSMO trilayer, 
as mentioned in the introduction we have also carried out model 
Hamiltonian simulations for the more standard case of a non-magnetic 
insulator as a barrier. A widely studied material for this purpose is 
STO, which is here mimicked simply by a tight-binding Hamiltonian for the 
barrier, with its energy levels shifted by a one-particle site potential 
$V_{\rm m}$ which controls the height of the barrier. 
In this section, 
results for the case LSMO/STO/LSMO are briefly discussed, with the emphasis 
on qualitative aspects and comparisons with the case of LMO as barrier. 
The details of the model Hamiltonian were already described in 
Sec.~\ref{model-methods}, and the technique employed is the MC 
simulation. Once again, it should be remarked that subtle 
effects such as ``dead layers'' are not considered in this study 
(Sec.~\ref{main_appr}), and their presence may affect quantitatively 
our conclusions.

\begin{figure}[hbt]
\includegraphics[clip=true,width=8.0cm,angle=-0]{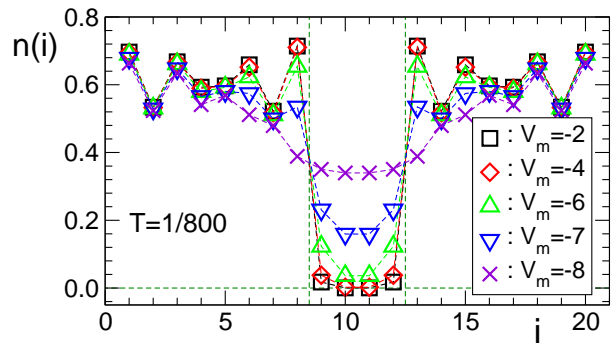}
\begin{center}
\caption{
(Color online) 
Electron density $n(i)$ for the 1D LSMO/STO/LSMO model (see the text) with 
$J_{\rm H}=8.0$ and $\alpha=1.0$, at $T=1/800$, and for various values of 
$V_{\rm m}$ (indicated in the figure). The positive background charges are 
$n_+^{(\rm L)}$=$n_+^{(\rm R)}$=$0.625$ and $n_+^{(\rm C)}$=$0.0$, and the 
system size studied is $L^{(\rm L)}$=$L^{(\rm R)}$=$8$ and 
$L^{(\rm C)}$=$4$. The positions of the two interfaces are 
denoted by vertical dashed lines. The electrostatic potential $\phi(i)$ is 
determined self-consistently. 
}
\label{den_BI}
\end{center}
\end{figure}

The local electronic density $n(i)$ is reported in Fig.~\ref{den_BI} for 
the 1D LSMO/STO/LSMO model with different values of the site potential 
$V_{\rm m}$ in the central region. It is observed that $n(i)$ in the 
central region gradually increases from nearly zero to $\sim0.4$ with 
decreasing $V_{\rm m}$, indicating that the barrier 
hight decreases with $V_{\rm m}$, and consequently the effective coupling 
between 
the leads becomes stronger. Note that the lower band of the lead (described 
by the DE model) is centered at $-J_{\rm H}$, and thus the band of the 
central region and the lower band of the lead are perfectly aligned 
(i.e., zero barrier height) when $V_{\rm m}=-J_{\rm H}$ (and $\alpha=0$).

\begin{figure}[hbt]
\includegraphics[clip=true,width=8.5cm,angle=-0]{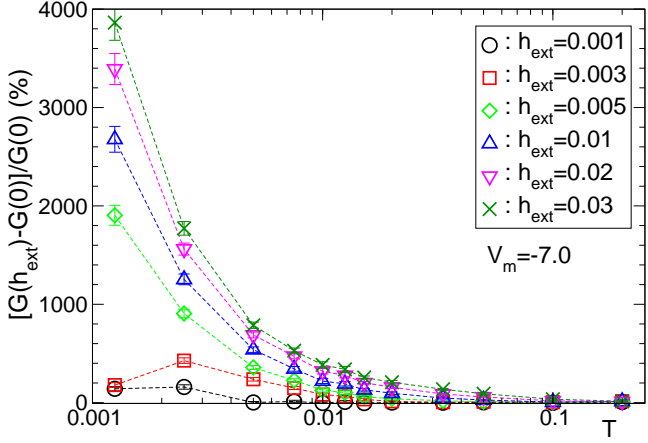}
\begin{center}
\caption{
(Color online) 
MR ratio $[G(h_{\rm ext})-G(0)]/G(0)\times100$ vs. temperature $T$ for the 
1D LSMO/STO/LSMO model (see the text) with $J_{\rm H}=8.0$, $\alpha=1.0$, 
and $V_{\rm m}=-7.0$. 
The positive background charges are $n_+^{(\rm L)}$=$n_+^{(\rm R)}$=$0.625$ 
and $n_+^{(\rm C)}$=$0.0$, and the system size studied is 
$L^{(\rm L)}$=$L^{(\rm R)}$=$8$ and $L^{(\rm C)}$=$4$.
Applied magnetic fields $h_{\rm ext}$ are indicated in the figure. 
Here, $\phi(i)$ is determined self-consistently for all temperatures. 
}
\label{TMR_V-7}
\end{center}
\end{figure}

\begin{figure}[hbt]
\includegraphics[clip=true,width=8.5cm,angle=-0]{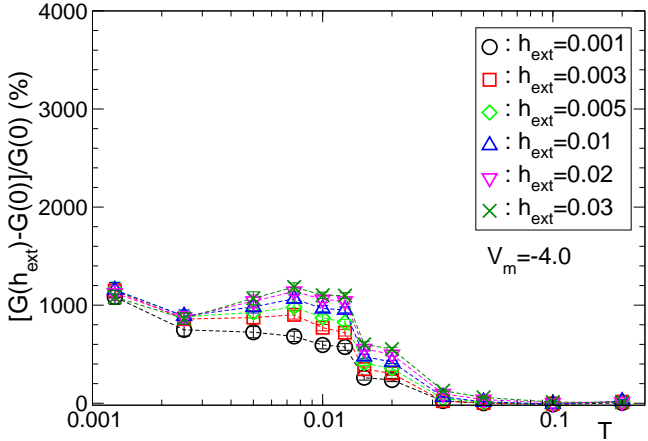}
\begin{center}
\caption{
(Color online) 
The same as in Fig.~\ref{TMR_V-7}, except that $V_{\rm m}$=$-4.0$.
}
\label{TMR_V-4}
\end{center}
\end{figure}

In Fig.~\ref{TMR_V-7}, the MR ratio vs. temperature 
for the case $V_{\rm m}=-7.0$ is shown. Notice that the MR effect becomes 
appreciable at temperatures comparable to the Curie temperature of the 
individual ferromagnetic leads. This is qualitatively similar to the 
effect observed 
in LSMO/LMO/LSMO for the case where an anisotropy was introduced 
(Fig.~\ref{TMR_s}), which effectively caused a stronger effective coupling 
between 
the magnetic moments of the left and right leads. Thus, it appears 
that the value $V_{\rm m}=-7.0$ chosen for this example allows for a robust 
left-right coupling 
(as also expected from $n(i)$ shown in Fig.~\ref{den_BI}), concomitant 
with the survival of a large MR effect up to the Curie temperature. 
It is also observed in Fig.~\ref{TMR_V-7} that because of the strong 
effective coupling between the leads, the MR ratio for small magnetic fields 
($\alt0.003$) is not as large as in the case of the small effective coupling 
discussed below (Fig.~\ref{TMR_V-4}).

Figure \ref{TMR_V-4} shows results for the same parameters as in 
Fig.~\ref{TMR_V-7}, but simply making the height of the barrier $V_{\rm m}$ 
much larger, i.e., the left and right leads being nearly decoupled. 
Three effects are obvious to the eye: (i) the large MR effect appears now at 
lower temperatures. For example, in Fig.~\ref{TMR_V-4} the MR becomes 
nonzero at T$\sim$0.02, while in Fig.~\ref{TMR_V-7}
the same occurred at a higher temperature $\sim0.03-0.04$. 
(ii) On the other hand, increasing the height of the barrier 
much increases the MR ratio at low temperatures and small magnetic 
fields. For example, in Fig.~\ref{TMR_V-4} the MR ratio is about 1,000\% 
at T$\sim$0.001 and $h_{\rm ext}=0.001$, while in Fig.~\ref{TMR_V-7} it is 
about 200\% at the same $T$ and $h_{\rm ext}$. This occurs because the barrier 
is so large in Fig.~\ref{TMR_V-4} that in the absence of magnetic fields, 
the magnetic moments of the left and right leads are almost decoupled, 
and thus applying a very small magnetic field is enough to align the leads' 
magnetic moment and increase conductance. 
(iii) The dependence of magnetic fields on the MR ratio is very mild 
(compared to Fig.~\ref{TMR_V-7}), namely 
changing $h_{\rm ext}$ from 0.001 to 0.03 only alters the results by a factor 
2 at the most. This is because the almost decoupled leads moments are forced 
to align in the magnetic field direction by a very small $h_{\rm ext}$. 
A further increase of $h_{\rm ext}$ does not change the 
moments orientation at low temperatures.

Thus, an interesting and simple picture emerges from these qualitative 
investigations, which are in agreement with the results shown before for 
the case of LSMO/LMO/LSMO: (1) If the barrier between the magnetic leads is 
very large (i.e., the effective coupling between the leads is small), the 
magnetic moments of the ferromagnetic leads are nearly decoupled 
(in the absence of anisotropies) at $h_{\rm ext}$=0, even at low 
temperatures. Thus, the double effect of large barrier and concomitant 
weakly coupled leads causes a large $h_{\rm ext}$=0 resistance, which 
enhances the MR effect at low temperatures as compared with a lower barrier 
(i.e., the large effective coupling between the leads). 
(2) However, the large barrier and weakly coupled leads are fragile upon 
increasing the temperature.

\section{Conclusions}\label{conclusions}

In this paper, the use of LSMO/LMO/LSMO as a spin-tunnel-junction is 
proposed. The main difference with other previous efforts is in the use of 
a manganite barrier. This would improve the lattice spacing matching between 
the constituents, hopefully also alleviating the complications found in 
previous investigations such as the infamous interfacial dead layers of 
STO/LSMO. However, the main point of this study is not just the all-manganite 
character of the trilayer, but the antiferromagnetic properties of LMO. 
For an even number of LMO layers, the spin order in the ground state is 
such that the magnetic moments of the ferromagnetic LSMO leads are 
anti-parallel (while for an odd number of layers, they are parallel). 
An anti-parallel-leads configuration has a large resistance. But the 
effective coupling leading to 
this anti-parallel LSMO-moments configuration is weak, rendering the ground 
state fragile. In fact, numerical simulations show that very small magnetic 
fields $h_{\rm ext}$ can alter drastically the original ground state at 
$h_{\rm ext}=0$, 
by aligning the magnetic moments of the ferromagnetic LSMO leads. A very 
large MR effect is observed in this transition, at least at low temperatures. 
Note that in arriving to our conclusions a large number of approximations 
have been made, all clearly described in Sec.~\ref{main_appr}. 
However, we still expect that our theoretical analysis is qualitatively 
correct and may serve as a motivation for a real experimental realization 
of the LSMO/LMO/LSMO magnetic junction.

Other effects, such as the influence of temperature, were also considered 
in this study. Together with the analysis of STO as barrier, overall trends 
were identified. Very insulating barriers (inducing a very week effective 
coupling between the leads) can lead to low-temperature states which are 
easily destabilized by small magnetic fields, causing a large MR effect. 
However, thermal fluctuation rapidly washes out these large effects. 
Reducing the barrier height or introducing anisotropies make the original 
ground state at $h_{\rm ext}=0$ more robust with increasing temperature, 
but this reduces the low-temperature MR effect at small magnetic fields. 
A balance between these two tendencies is needed to find optimal trilayers 
for real devices.

{\it Note added:} While completing the work described in this paper, we 
received two preprints with interesting related efforts: 
(1) Salafranca {\it et al.} \cite{salafranca} have reported a theoretical 
study of an all-manganite heterostructure consisting of FM electrodes and 
an AF barrier, similar in spirit to ours. 
However, contrary to our proposed system, 
the chosen barrier~\cite{salafranca} is Pr$_{2/3}$Ca$_{1/3}$MnO$_3$, 
which has CE-type ordering and the same hole doping as the electrodes, which 
were chosen to be LSMO with $x$=1/3. The emphasis of 
Ref.~\onlinecite{salafranca} is not on differences between even and odd 
number of central layers as in the present paper, but on other 
interesting effects such as the influence of the FM electrodes on 
the spin arrangement of the barrier. Thus, Ref.~\onlinecite{salafranca} 
and our efforts nicely complement each other. 
(2) Yu {\it et al.}\cite{yu} have presented experimental results for an 
all-manganite trilayer using LSMO $x$=0.3 as electrodes and LSMO $x$=0.04 
as barrier (the latter being almost identical to the LMO barrier 
theoretically proposed in this paper). Those authors report a huge TMR ratio 
of 30,000\% at 4.2~K and with bias voltage 25~mV (our MR ratio is 200,000\% 
at $h_{\rm ext}$=0.001). The barrier thickness is 9 atomic layers. 
Yu {\it et al.} assign the large MR effect they observed to thermally 
activated magnon resonances inside the barrier. A detail comparison 
between theory and experiment will  be carried out in the near future.

\begin{acknowledgments}

We thank Hiroshi Akoh, Luis Brey, Satoshi Okamoto, 
Hiroshi Sato, and Hiroyuki Yamada for very useful discussions,
and X.-G. Zhang for bringing Ref.~\onlinecite{yu} to our attention.
This work was supported in part by the NSF grant
DMR-0706020 and by the Division of Materials Science and Engineering, 
U.S. DOE, under
contract with UT-Battelle, LLC.

\end{acknowledgments}


\end{document}